\documentclass[10pt,aps,prd,twocolumn,notitlepage,nofootinbib,floatfix,
               superscriptaddress]{revtex4-1}

\usepackage{amsmath, amsthm, amssymb, amsfonts, amsbsy, mathrsfs}

\usepackage{xcolor}
\usepackage{calligra,bm}
\usepackage{stmaryrd}

\usepackage{graphicx}
\graphicspath{{./images/}}

\usepackage[hidelinks,bookmarks=true]{hyperref} 
\hypersetup{pdfstartview=FitH,pdfhighlight=/O,colorlinks=false}

\begin{document}

\title{Smooth metrics can hide thin shells}

\author{Justin C. Feng}
\affiliation{CENTRA, Departamento de F{\'i}sica, Instituto Superior
             T{\'e}cnico – IST, Universidade de Lisboa – UL, Avenida
             Rovisco Pais 1, 1049 Lisboa, Portugal}
\affiliation{Center for Gravitational Physics, The University of Texas
at Austin, Austin, TX 78712, USA}

%
%

%
%
\begin{abstract}
    In this note, I consider a class of metric tensors with smooth
    components that naively appear to describe dynamical wormholes with
    vanishing spacetime curvature. I point out that the smoothness of
    the metric tensor components is deceptive, and that in general
    relativity, such metrics must be sourced by exotic thin shells.
\end{abstract}

%
%

\maketitle

%
%
%

This note is motivated by the dynamical generalization of a flat
``defect wormhole'' metric that appears in
\cite{Klinkhamer:2022rsj,*Klinkhamer:2023sau,*Klinkhamer:2023avf,*Klinkhamer:2023nok},
which has attracted some attention lately
\cite{Wang:2023rfq,Ahmed:2023dvc}. The construction of the static metric
is rather straightforward. Consider the line element for flat spacetime,
written in spherical coordinates (I employ the $(-,+,+,+)$ signature
throughout):
\begin{equation}\label{Eq-FlatLE}
    ds^2 = - dt^2 + dr^2 + r^2 d^2\Omega,
\end{equation}
where $d^2\Omega:=d\theta^2+\sin^2\theta d\phi$ is the standard line
element on the unit 2-sphere. Now perform a coordinate transformation
defined by the relation:
\begin{equation}\label{Eq-CoordTransform}
    r = \sqrt{\lambda^2+\rho^2},
\end{equation}
where $\lambda$ is assumed to be a positive real parameter and $\rho \in
\{-\infty,\infty\}$. One obtains the line element
\cite{Klinkhamer:2022rsj,Klinkhamer:2023sau}:
\begin{equation}\label{Eq-WWLE}
    ds^2 = - dt^2 + \frac{\rho^2}{\lambda^2+\rho^2} d\rho^2 + 
            (\lambda^2+\rho^2) d^2\Omega,
\end{equation}
which describes a spacetime with topology $R \times R \times S^2$ and
two asymptotically flat regions, one in the limit
$\rho \rightarrow \infty$, and another in the limit
$\rho \rightarrow -\infty$. A direct evaluation for generic $\rho$
yields a vanishing Riemann tensor, so that one might naively interpret
Eq. \eqref{Eq-WWLE} as describing a flat wormhole with a throat at
$\rho=0$.

Upon adding a time dependence $\lambda\rightarrow\lambda(t)$ in Eq.
\eqref{Eq-CoordTransform}, one obtains the dynamical generalization of
Eq. \eqref{Eq-WWLE}:
\begin{equation}\label{Eq-WWLETD}
    \begin{aligned}
    ds^2 =\, & 
     -  \left[
        1-\frac{[\lambda(t) \lambda^\prime(t)]^2}{\lambda^2+\rho^2}
        \right]
        dt^2 
    + \frac{2\lambda(t) \lambda^\prime(t)}{\lambda(t)^2+\rho^2} dt d\rho
    \\
    &+ \frac{\rho^2}{\lambda(t)^2+\rho^2} d\rho^2 + 
            \left[\lambda(t)^2+\rho^2\right] d^2\Omega.
    \end{aligned}
\end{equation}
Since the Riemann tensor vanishes for generic $\rho$, one might naively
imagine that the geometry described by Eq. \eqref{Eq-WWLETD} is
everywhere flat, but if that were the case, the function
$\lambda(t)$ (the areal radius of the throat $\rho=0$) would be
unconstrained by the Einstein equation. Such a feature would be
disastrous for general relativity, rendering the initial value problem
ill-posed, in conflict with accepted theorems, for instance
Thm. 10.2.2 of \cite{Wald}, which establishes that given smooth initial
data, there exists a unique spacetime that is a solution to the Einstein
equation and is continuously dependent on the initial data.

So what gives? It turns out that the smoothness of the metric tensor
components is deceptive---a closer inspection of Eq. \eqref{Eq-WWLE} at
$\rho=0$ reveals hints of the deception. In particular, $r(\rho)$ in Eq.
\eqref{Eq-CoordTransform} is not bijective in a
neighborhood of $\rho=0$ containing both positive and negative values of
$\rho$, and the $d\rho^2$ term in Eq. \eqref{Eq-WWLE} vanishes at
$\rho=0$, rendering the metric degenerate there, as pointed out in
\cite{Klinkhamer:2022rsj,*Klinkhamer:2023sau,*Klinkhamer:2023avf}
(though degenerate metrics can in some cases indicate coordinate
singularities, at $\theta=0$ for instance). A more suggestive hint
comes from the observation \cite{MukohyamaPC} that the expansion scalar
for a null radial geodesic congruence changes sign as the congruence
passes through the throat $\rho=0$; the same property is used
to show that spherically symmetric wormholes violate energy conditions
(see \cite{Morris:1988cz} and Sec. 13.4.2 of \cite{Visser1995}). Indeed,
the mystery unravels when explicitly computing limits of the extrinsic
curvature tensor $K_{\mu\nu}:=\tfrac{1}{2} \pounds_n (g_{\mu\nu}-n_{\mu}
n_{\nu})$ for constant $\rho$ surfaces (which are timelike), where
$\pounds_n$ is the Lie derivative with respect to $n^{\mu}$, the unit
normal vector to a surface of constant $\rho$:
\begin{equation}\label{Eq-ExtrinsicCurvatureLimits}
    \begin{aligned}
        \lim_{\rho_+\rightarrow0}K_{\mu \nu} 
        &= \text{diag}(0,0,\lambda,\lambda\sin^2\theta) \\
        \lim_{\rho_-\rightarrow0}K_{\mu \nu} 
        &= \text{diag}(0,0,-\lambda,-\lambda\sin^2\theta) .
    \end{aligned}
\end{equation}
$\rho_+\rightarrow0$ indicates the limit for $\rho>0$, and
$\rho_-\rightarrow0$ indicates the limit for $\rho<0$. The extrinsic
curvature tensor for the family of constant $\rho$ surfaces is
discontinuous in $\rho$ at $\rho=0$, signaling the presence of a
distributional contribution to the curvature (see \cite{Israel1966}
and Sec. 3.7 of \cite{Poisson}). The Einstein tensor acquires a
nonvanishing term proportional to $\delta(\rho)$, and it follows from
the Einstein equation that the line element \eqref{Eq-WWLE} is sourced
by a (necessarily exotic) thin shell at $\rho=0$.

One can draw two conclusions from this result. The first is that the
line elements \eqref{Eq-WWLE} and \eqref{Eq-WWLETD} describe thin shell
wormholes \cite{Visser1989,Visser1995}, which have been studied
extensively in classical general relativity (see
\cite{Alcubierre:2017pqm} and references therein) and as toy models in
quantum gravity \cite{Redmount:1992mc,*Redmount:1993ue}. The second is
the eponymous conclusion that metric tensors with
smooth components can hide thin shell sources---one should check for
the presence of thin shells at hypersurfaces on which a metric becomes
degenerate.


{\em Acknowledgements --- I thank S. Mukohyama, R. A. Matzner,
and F. R. Klinkhamer for clarifying remarks and feedback.}


%
%


\bibliography{ref}

\end{document}